\documentclass[11pt, draftcls, a4paper, onecolumn]{IEEEtran}

\usepackage{amsmath,epsfig,amssymb,verbatim,amsopn,cite,multirow}
\usepackage{balance,nopageno}
\usepackage{algorithm,algorithmic}
\usepackage[usenames,dvipsnames]{color}
\usepackage[all]{xy}  
\usepackage{url}
\usepackage{amsfonts}
\usepackage{amssymb}
\usepackage{epsfig}
\usepackage{epstopdf}
\usepackage{slashbox}
\usepackage{bm}
\usepackage{balance}
\usepackage{graphicx}

\usepackage{subfig}
\usepackage[margin=21.0mm,top=20mm,bottom=20mm]{geometry}

\newcommand{\Nrx}{\mathsf{N_{R}}}
\newcommand{\Ntx}{\mathsf{N_{T}}}
\newcommand{\Nsx}{\mathsf{N_{S}}}

\newcommand{\DL}{\text{DL1}}
\newcommand{\DLL}{\text{DL2}}
\newcommand{\UL}{\text{UL1}}
\newcommand{\ULL}{\text{UL2}}
\newcommand{\CU}{\text{CU1}}
\newcommand{\CUU}{\text{CU2}}
\newcommand{\Ith}{I_\mathsf{th}}

\title{Full-Duplex Non-Orthogonal Multiple Access for Modern Wireless Networks}
\author{\normalsize {Mohammadali Mohammadi,~\IEEEmembership{Member,~IEEE,}
Xiaoyan Shi,~\IEEEmembership{Student Member,~IEEE,}\\
Batu K. Chalise,~\IEEEmembership{Senior Member,~IEEE,}
Himal A. Suraweera,~\IEEEmembership{Senior Member,~IEEE,}\\
\hspace{1.25em}Caijun Zhong,~\IEEEmembership{Senior Member,~IEEE,}
 and John S. Thompson,~\IEEEmembership{Fellow,~IEEE}}
  \thanks{
    M. Mohammadi is with the Faculty of Engineering, Shahrekord University, Shahrekord 115, Iran
    (email: m.a.mohammadi@eng.sku.ac.ir).}
    \thanks{
    B. K. Chalise is with the Cleveland State University, 2121 Euclid Avenue, Cleveland, OH 44115, USA (email: b.chalise@csuohio.edu). }
  \thanks{
    H. A. Suraweera is with the Department of Electrical and Electronic
Engineering, University of Peradeniya, Peradeniya 20400, Sri Lanka  (email:
    himal@ee.pdn.ac.lk). }
      \thanks{
    C. Zhong is with the Department of Information Science and Electronic
Engineering, Zhejiang University, Hangzhou 310027, China (email:  caijunzhong@zju.edu.cn). }
      \thanks{
  X. Shi and J. S. Thompson are with the Institute for Digital Communications, School
of Engineering, University of Edinburgh, United Kingdom (email: {Xiaoyan.Shi, John.Thompson}@ed.ac.uk). }

}

\begin{document}
\maketitle
\begin{abstract}
Non-orthogonal multiple access (NOMA) is an interesting concept to provide higher capacity for future wireless communications. In this article, we consider the feasibility and benefits of combining full-duplex operation with NOMA for modern communication systems. Specifically, we provide a comprehensive overview on application of full-duplex NOMA in cellular networks, cooperative  and cognitive radio networks, and characterize gains possible due to full-duplex operation. Accordingly, we discuss challenges, particularly the self-interference and inter-user interference and provide potential solutions to interference mitigation and quality-of-service provision based on beamforming, power control, and link scheduling. We further discuss future research challenges and interesting directions to pursue to bring full-duplex NOMA into maturity and use in practice.
\end{abstract}
\section{Introduction}
The proliferation of multimedia services, coupled with ever growing number of mobile subscribers, has led to a demand for increased spectral efficiency in the emerging wireless communication networks such as fifth generation (5G) cellular communication networks. To meet this demand, full-duplex communication, i.e., simultaneous transmission and reception  on the same channel at the same time, has the potential to double the spectral efficiency of traditional half-duplex networks~\cite{Ashutosh:JSAC:2014}. In practice, however, the increase in throughput due to full-duplex operation is limited by the presence of unavoidable self-interference (SI) caused by the signal leakage from the transceiver output to the input~\cite{Riihonen:JSP:2011}. Recent advances in antenna and transceiver design enable SI cancellation at practical costs, and managed to limit SI up to the receiver noise floor~\cite{Sabharwal:TWC2012}. Therefore, full-duplex technology has been recognized as a promising candidate for 5G network design.

On a parallel development, non-orthogonal multiple access (NOMA) is another promising concept for improving the spectral efficiency of next generation wireless networks~\cite{Saito:VTC2013}. As such a downlink version of NOMA, termed as multiuser superposition transmission (MUST), has been included in the 3GPP long term evolution-advanced (LTE-A) standard~\cite{Ding:Survay}. NOMA leverages its flexible architecture to provide coverage and throughput expansion in a cost effective manner.
Although user multiplexing through power allocation among the users reduces the allocated power to each single user, both users with strong  channel condition (which we refer to as NOMA-strong user) and users with poor channel conditions (which we refer to as NOMA-weak user) benefit from being scheduled more often and being assigned more bandwidth. Therefore, the average throughput can be increased significantly at a moderate increase of the system complexity~\cite{Saito:VTC2013,Ding:Survay}. Moreover, NOMA is compatible with orthogonal frequency division multiple access in the downlink, and thus the number of simultaneously served users can be almost doubled (up to two users multiplexed in power-domain at the same sub-band), which is suitable to address the challenges related to massive connectivity and low transmission latency in 5G wireless networks. Furthermore, NOMA provides robust performance even in high mobility scenarios, since the NOMA transmitter does not rely much on channel state information (CSI) feedback from the receiver for user multiplexing~\cite{Ding:TVT:2016}. NOMA can be also applied to multi-antenna networks which provide additional degrees-of-freedom for further performance improvement~\cite{Zhiguo-MIMO}. The multi-antenna aspect has also been addressed in the context of MUST, showing that multiple-input-multiple-output (MIMO) technology and NOMA can be used jointly to provide combined gain.

Most of the work on NOMA to date, however, is limited to the half-duplex operation. Since both NOMA and full-duplex techniques improve the spectral efficiency, it would be interesting to investigate the additional benefit that can be achieved by combining them.
A potential application of full-duplex transceivers in NOMA systems is to enable simultaneous uplink and downlink transmissions in a cellular network, where data from the paired users in the uplink channel, and data to the paired users in the downlink channel are transmitted and received at the same time on the same frequency. Furthermore, by incorporating the full-duplex operation in downlink transmissions, a NOMA user close to the base station (BS)  can receive data and simultaneously forward data to a more distant NOMA user over the same carrier frequency. Hence, the requirement for extra time slots for relaying is avoided. There have been few recent studies on the combination of full-duplex operation and the NOMA principle, which unanimously concede that the consideration of full-duplex communications in the NOMA-based networks is not a trivial task. The main concern with full-duplex operation is that transmissions suffer from both SI and inter-user interference, which significantly affect the design process of NOMA systems and degrade the system performance. In this context, we clearly show how SI and inter-user interference affect full-duplex NOMA systems using several systems as examples: uplink and downlink cellular communications, cooperative, and cognitive radio networks.

In this article, we first present the state-of-the-art on full-duplex NOMA systems.  Then, we introduce various full-duplex NOMA designs for several networks and quantify the performance gain due to full-duplex transmissions. Specifically, we characterize the uplink and downlink full-duplex NOMA, where a full-duplex BS serves uplink and downlink half-duplex users simultaneously in the same radio resource. The concept of cooperative full-duplex NOMA will then be described and consequently the application of the NOMA in the full-duplex cognitive radio networks will be discussed.  From the optimization algorithm perspective, we also discuss resource management for a NOMA system operating with the full-duplex technique. Finally, research challenges and some promising future directions for designing full-duplex NOMA systems are highlighted.
\section{Full-duplex NOMA Transmissions}
\subsection{NOMA Basics}
NOMA realizes multiple access by exploiting the power domain, which is fundamentally different from conventional orthogonal
multiple access (OMA) technologies~\cite{Ding:Survay,Zhiguo-MIMO}. In contrast to the  conventional ``water-filling" power allocation, NOMA allocates more power to users with worse channel conditions than those with better channel conditions, thus, imposing an additional fairness constraint.

For the sake of conceptual clarity, let us consider a single cell downlink NOMA with one BS and two users, denoted as $\DL$ and $\DLL$ respectively. Consider that $\DLL$ is located close to the cell-edge, and hence suffering from poor channel conditions. BS transmits a superposition coded signal, which is a sum of both users' signals. According to the NOMA principle, the transmit power of the information signal for $\DLL$ must be greater than that for $\DL$. Since $\DLL$'s channel condition is very poor, the interference
from $\DL$, caused by the superimposed transmit signals, will not cause much performance degradation to $\DLL$.  Accordingly, $\DLL$ decodes directly its information signal by treating the interference induced by the $\DL$ as noise. In contrast, $\DL$ can decode its own information signal
after removing the $\DLL$'s signal through a successive interference cancellation (SIC) receiver. Therefore, both $\DL$ and $\DLL$ are scheduled for communication at a time-slot, hence, the system throughput can be significantly improved compared to the OMA schemes.
\subsection{State-of-the-Art on Full-duplex NOMA Systems}
The possible use cases of above described NOMA concept benefiting from full-duplex operation are a full-duplex BS serving uplink and downlink users simultaneously in the same radio resource~\cite{FD-NOMA}, and cooperative NOMA systems where full-duplex NOMA users~\cite{Z.Zhang,Nallanathan:FDNOMA} or dedicated full-duplex relays~\cite{Caijun:CLET:2016} assist the transmissions between the source and NOMA-weak users. The authors of~\cite{FD-NOMA} investigated the resource allocation algorithm design for a full-duplex multicarrier NOMA system, where a full-duplex BS is simultaneously serving multiple half-duplex downlink and uplink users.
The entire frequency band is partitioned into multiple orthogonal subcarriers, where each subcarrier is allocated to at most two downlink users and two uplink users. Resource allocation algorithm design for the maximization of the weighted sum throughput of the system is formulated as a non-convex optimization problem and the optimal power and subcarrier allocation policies have been obtained.

The second case is termed as full-duplex cooperative NOMA, where different users in a NOMA network cooperate with each other to enhance the performance. Cooperative relaying has been extensively studied as an effective method for establishing connectivity between the nodes of the network. Hence, cooperative NOMA is a natural extension of relaying systems that can take advantage of the reduced attenuation between the relay node and NOMA-weak users~\cite{Z.Zhang,Nallanathan:FDNOMA,Caijun:CLET:2016,Jungho:CLET:2016}. In~\cite{Z.Zhang}, a full-duplex device-to-device aided cooperative NOMA scheme was proposed where the NOMA-strong user is full-duplex capable. Moreover, the authors in~\cite{Z.Zhang} proposed an adaptive multiple access scheme, which dynamically switches between the proposed cooperative NOMA, conventional NOMA, and OMA schemes, according to the level of residual SI and the quality of links, and outperforms all other multiple access schemes. The authors in~\cite{Nallanathan:FDNOMA} provided a diversity analysis for cooperative full-duplex NOMA systems and proved that the use of the direct link overcomes the lack of diversity for the NOMA-weak user which is otherwise inherent to the full-duplex relaying. In~\cite{Caijun:CLET:2016} a dual-user NOMA system was studied, where a dedicated full-duplex relay assists information transmission to the user with the weaker channel condition. The proposed full-duplex cooperative NOMA system~\cite{Caijun:CLET:2016} achieves higher ergodic sum capacity compared to the half-duplex cooperative NOMA counterpart in the low to moderate signal-to-noise (SNR) regimes. Methods to determine possible data rates for a full-duplex cooperative NOMA system with relaying were proposed in~\cite{Jungho:CLET:2016}.

\begin{figure}[h]
\centering
\vspace{-20em}
\includegraphics[scale=0.50]{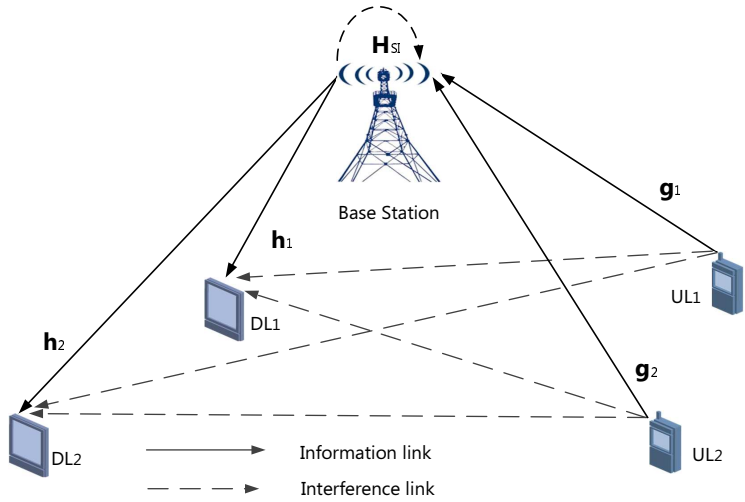}
\vspace{0em}
\caption{System model: Full-duplex NOMA with two downlink users and two uplink users. }\label{fig:SM_UD}
\vspace{-1em}
\end{figure}

Most of the current work on full-duplex NOMA shows that the achievable gains are sensitive to the amount of residual SI power. In practice, the residual SI bring new challenges in making a SIC receiver feasible. Each full-duplex receiver needs to first cancel its own SI, and then proceed to decode the inter-user interference signals and finally recover its own signal. Therefore, the accuracy of the adopted SI canceller will play a pivotal role in the SIC process. The increased interference at user terminals in full-duplex scenarios will make it difficult for the receiver to cancel the strong signal before decoding its own message. In order to fully exploit the benefits of full-duplex NOMA, advanced SI cancellation techniques and approaches such as coordinated multi-point, interference alignment  coupled with sophisticated error correction coding schemes appear to be necessary.
\section{Uplink and Downlink Full-duplex NOMA}
As cellular networks evolve, there has been an increasing emphasis on the need for efficient transmission schemes. In legacy cellular networks, orthogonal resource allocation mechanisms limit the maximum number of supported users with the finite resources.
NOMA can accommodate multiple users simultaneously via non-orthogonal resource allocation.
\begin{figure}[htb!]
\centering
\includegraphics[scale=0.55]{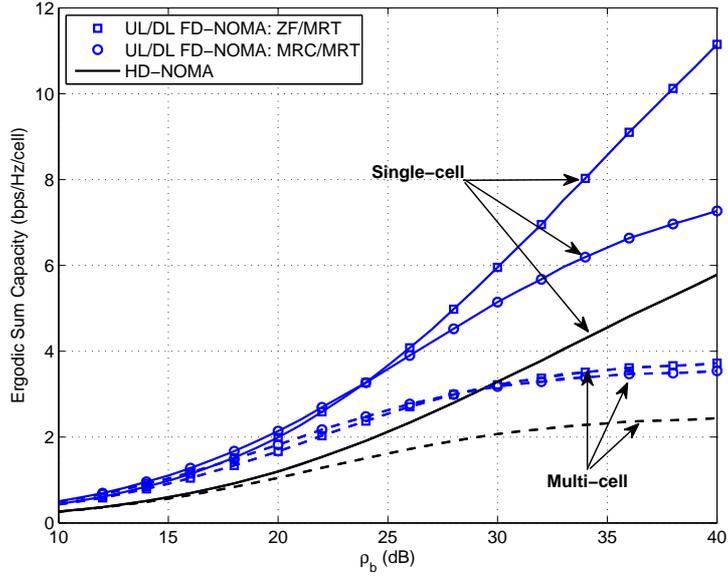}
\caption{Ergodic capacity comparison between the uplink/downlink full-duplex and half-duplex NOMA systems in single-cell and multi-cell scenarios and for $\Ntx = 3$ transmit antennas and $\Nrx=2$ receive antennas at the BS and for $\sigma^2_{\mathsf{SI}}=-10$ dBW.}\label{fig:uldl}
\end{figure}
Initial system implementations of NOMA in cellular networks have demonstrated the superior spectral efficiency of NOMA.

A potential way to achieve higher spectral efficiency in NOMA-based cellular networks is through the use of full-duplex transmission at the BS, so that the BS can transmit to multiple paired downlink users and receive from multiple paired uplink users in the same time and frequency band. However, the benefits of full-duplex communication for cellular networks cannot be reaped without having some cost increment. As an illustrating example, Fig.~\ref{fig:SM_UD} shows a simple dual-user system where $\DL$ ($\UL$) and $\DLL$ ($\ULL$) represent a NOMA-strong downlink (uplink) user and a NOMA-weak downlink (uplink) user, respectively. As illustrated in Fig.~\ref{fig:SM_UD}, the uplink communication is affected by the SI at the BS and the downlink communication suffers from the inter-user interference resulting from the uplink users sharing the same time/frequency resource.

The application of multiple antennas to NOMA provides additional degrees-of-freedom for further performance improvement. A transmission framework based on signal alignment was proposed in~\cite{Zhiguo-MIMO} for MIMO-NOMA systems, which is applicable to both uplink and downlink transmissions and offers a significant performance gain in terms of reception reliability. However, full-duplex cellular networks differ from their half-duplex counterparts in that the uplink transmission will be affected by the SI and the uplink users will interfere with the downlink reception. Therefore, in practice, the transceiver design for full-duplex BS must be revisited to achieve the promised gain of the full-duplex operation. The BS could use its multiple antennas for spatial SI cancellation and consequently for improving the signal-to-interference-plus-noise ratio (SINR) of both uplink and downlink transmissions.
In Fig.~\ref{fig:uldl} we present a numerical example demonstrating the ergodic sum capacity for multi-cell and single-cell environments. Simulations adopt parameters of the 3GPP LTE for small cell deployments. The receive beamformer at the BS is designed with a zero-forcing (ZF) constraint. Transmit beamformer is designed based on the maximum ratio transmission (MRT) principle.  In single-cell environment the half-duplex system outperforms the full-duplex system with maximum ratio combining (MRC)/MRT scheme at high SNR. The results in multi-cell scenario demonstrate that full-duplex operation significantly outperforms the half-duplex counterpart over the entire SNR regime.
\begin{figure}[htb]
\centering
\subfloat[HD: User assisted]{\includegraphics[width=.45\textwidth]{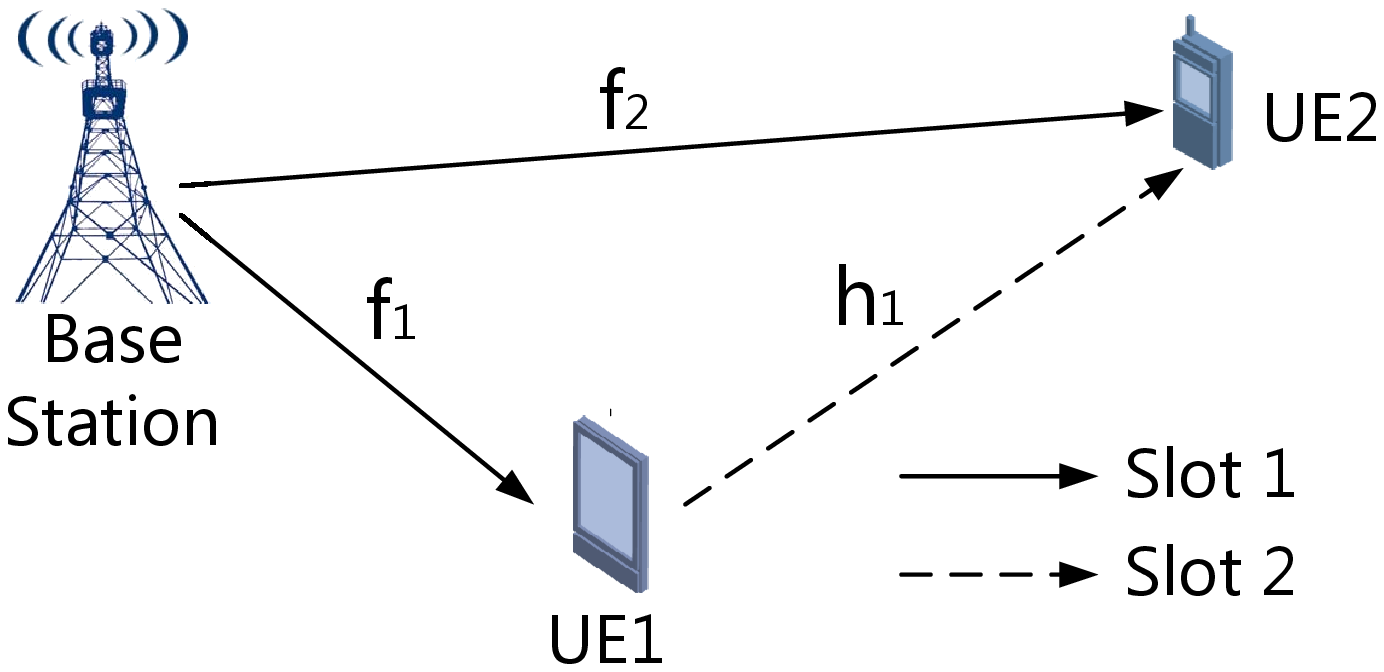}}\hspace{0.5cm}
\subfloat[HD: Relay assisted]{\includegraphics[width=.45\textwidth]{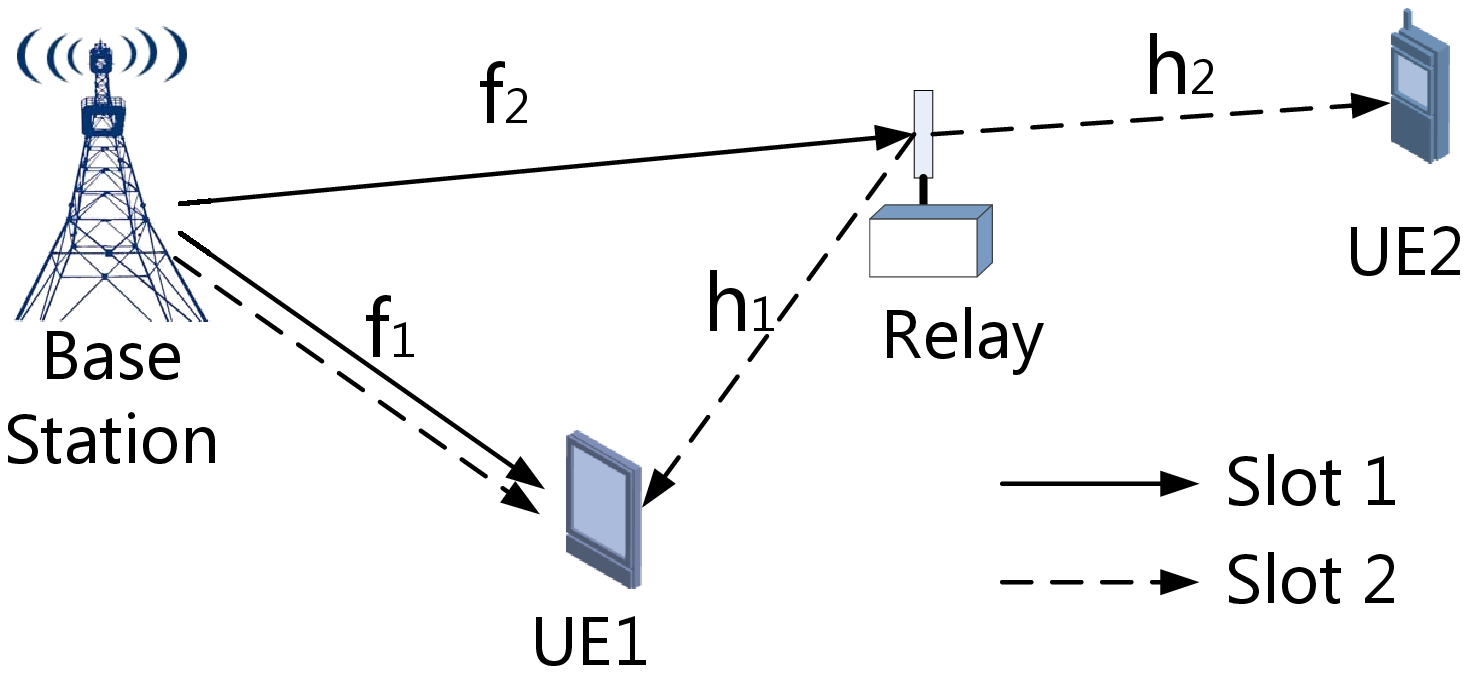}}\\
\subfloat[FD: User assisted]{\includegraphics[width=.45\textwidth]{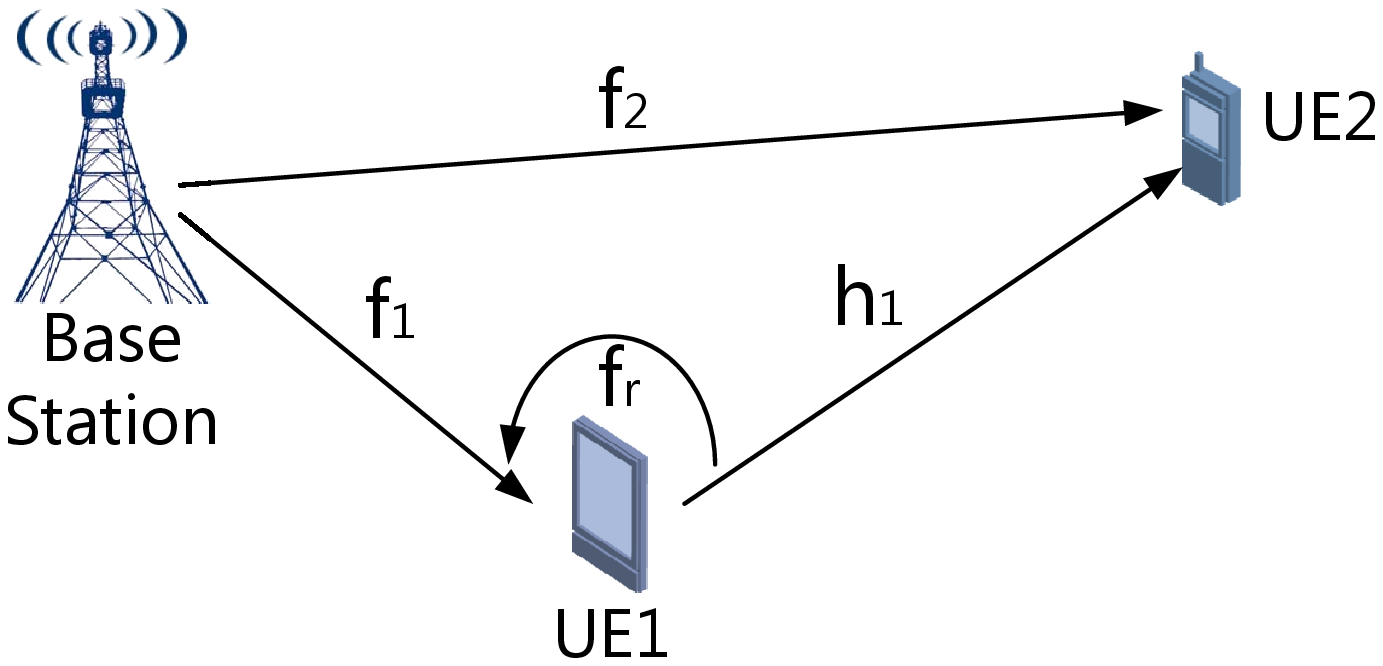}}\hspace{0.5cm}
\subfloat[FD: Relay assisted]{\includegraphics[width=.45\textwidth]{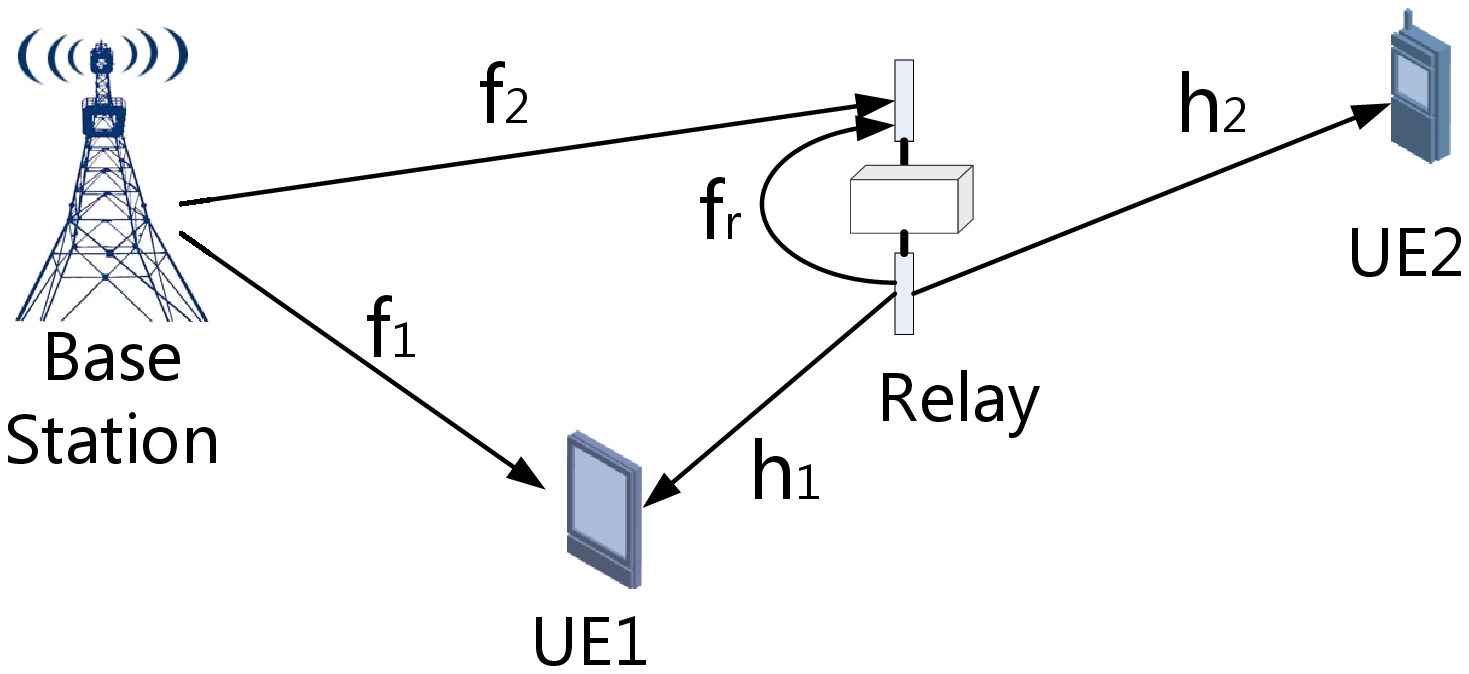}}
\caption{System model: Dual-user cooperative NOMA.}
\label{fig:1}
\end{figure}
Full-duplex NOMA system achieves $75\%$ higher average sum rate than the half-duplex counterpart at SNR of $20$ dB. In a multi-cell environment, co-channel interference will determine the performance gap between full-duplex and half-duplex modes. However, co-channel interference can be reduced significantly by exploiting techniques such as interference coordination.  Moreover, choice of beamforming design will also be crucial to realize the benefits of full-duplex NOMA systems as  requirements of the near/far users as well as constraints imposed by the NOMA principle must be incorporated.
\section{Cooperative Full-duplex NOMA}
It is well-known that cooperative communications can be used to extend the coverage and
improve the communication reliability. The attractive features of cooperative communication can be reaped using relatively simple signal processing techniques, however its potential is not yet fully explored for NOMA transmissions. In the literature, there are two different types of cooperative NOMA systems, namely, the user-assisted cooperative NOMA \cite{Z.Ding} and the relay-assisted cooperative NOMA \cite{J.Kim}. To explain the basic concept, let us consider the dual-user cooperative NOMA system as illustrated in Fig.~\ref{fig:1}(a) and~\ref{fig:1}(b). For user-assisted cooperative NOMA, the NOMA-strong user, helps the NOMA-weak user, exploiting the fact that the NOMA-strong user is able to decode the information for both users. For the relay-assisted cooperative NOMA, a dedicated relay is employed to assist the NOMA-weak user.

\begin{figure}[htb!]
\centering
\includegraphics[scale=0.7]{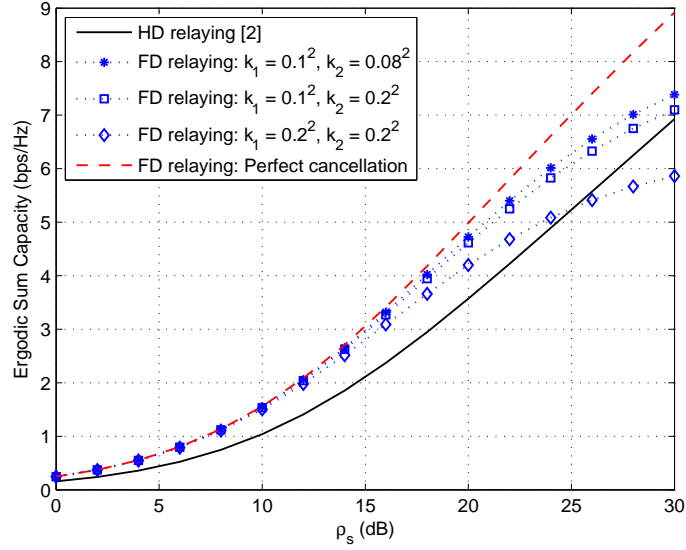}
\caption{Ergodic capacity comparison between the relay-assisted half-duplex and full-duplex cooperative NOMA systems with relay transmit power $\rho_r = \rho_b/2$, power allocation coefficients $a_1 = 0.05$ and $a_2 = 0.95$,  channel gains for $f_1$ being $\lambda_{b1} = 1$, for $f_2$, $h_1$ and $h_2$ being $\lambda_{br} = \lambda_{r1}=\lambda_{r2} = 0.5$, while for $f_r$ being $\lambda_{rr} = 0.3$.}\label{fig:2}
\end{figure}
While cooperation improves the reliability of the NOMA systems, the implementation of cooperation among users or relays requires additional time resources, which results in a loss of spectral efficiency. Since one of the key features of NOMA is the improvement of spectral efficiency, such spectral efficiency degradation due to cooperation is highly undesirable. To tackle this critical issue, and capitalize on the recent advances in full-duplex  communications, we propose to empower the cooperative node with the full-duplex capability as illustrated in Fig.~\ref{fig:1}(c) and~\ref{fig:1}(d), where both the NOMA-strong user and relay operate in the full-duplex mode. In half-duplex relays, two time slots are required to forward data to a relay and then to the destination. The full-duplex cooperative systems eliminate the additional time slot required for cooperation, which potentially doubles the spectral efficiency. However, the adoption of full-duplex may significantly elevate the interference level. For an example, consider the relay-assisted full-duplex cooperative NOMA, which, in addition to the SI at the full-duplex relay also creates harmful interference at the NOMA-strong user. To see the practical gain of full-duplex cooperative NOMA, let us consider the simple model shown in Fig.~\ref{fig:1}(d), where the BS transmits the superimposed signal to user equipment UE1 and the relay, while the relay overhears the superimposed signal, and at the same time, transmits a previously received signal intended for UE2. Since both UE1 and relay are aware of the signal intended for UE2, known interference cancellation techniques can be applied. However, due to imperfect cancellation, residual interference is likely to exist. Let $k_1$ and $k_2$ denote the residual interference power level at UE1 and the relay, respectively.
\begin{figure}[htb]
\vspace{-12em}
\centering
\subfloat[System model (Dashed lines denote the interference links.)]
{\includegraphics[width=.48\textwidth]{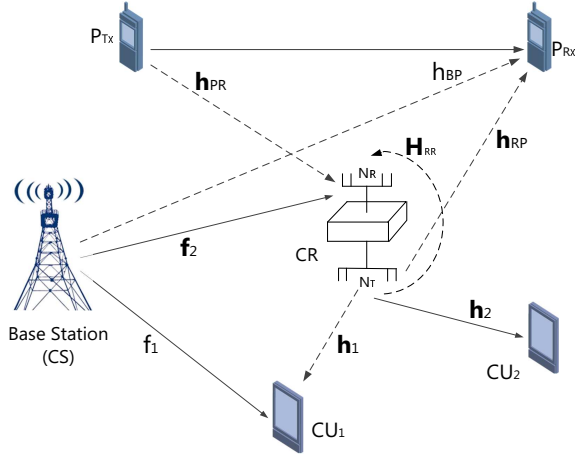}}\hspace{0.5cm}
\subfloat[Rate region achieved by the optimum and the suboptimum schemes.  ]{\includegraphics[width=.48\textwidth]{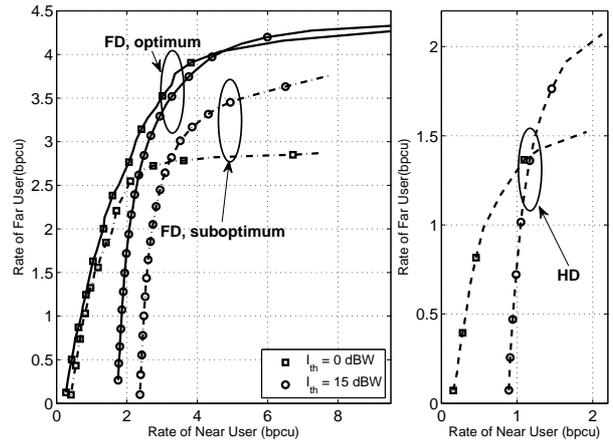}}\\
\caption{Cognitive NOMA with full-duplex relaying.}
\label{fig:3}
\end{figure}
According to Fig.~\ref{fig:2}, the ergodic sum capacity of the full-duplex cooperative NOMA system is better than that of the half-duplex cooperative NOMA system in the low to moderate SNR range. Moreover, the gain is less sensitive to the strength of the residual SI power, especially in the low SNR regime. This important observation indicates that the gain maybe realized without stringent requirements on SI cancellation capability. In contrast, the ergodic sum capacity of the full-duplex cooperative NOMA system is worse than that of the half-duplex  cooperative NOMA system in the high SNR regime, mainly due to high interference level, indicating the critical importance of SI mitigation in this regime.
\section{ NOMA in Full-duplex Cognitive Radio Networks}
NOMA can be viewed as an special case of cognitive radio networks (CRNs), where a user with a strong channel
condition is squeezed into the spectrum occupied by a user with a poor channel condition. The strong-user can be viewed as a primary user and the weak-user can be viewed as cognitive user in the CRN. Therefore, according to the principle of CRNs, the transmit power allocated to the strong-user is constrained by the weak-user's SINR. Following this concept, a variation of NOMA, termed cognitive radio inspired NOMA has been proposed in~\cite{Ding:TVT:2016}.

The NOMA concept is applicable for underlay CRNs, where a cognitive source (CS) communicates with two or more cognitive users (CUs) using the NOMA concept. All transmissions of the cognitive network occur in the transmission band of the primary network, provided any interference caused to a primary user is kept below a predefined threshold. Given this interference constraint, the performance of the cognitive network is strictly limited. Cooperative techniques help to reduce the transmission power as the communication distances in cognitive network are decreased by using a cognitive relay (CR). Hence, the mutual interference between primary and cognitive networks can be managed effectively. If the CR operates in a half-duplex mode, at least two orthogonal communication phases are needed, which reduces the throughput. This loss in spectral efficiency can be mitigated with the help of full-duplex radio at the CR as shown in Fig.~\ref{fig:3}(a). On the other hand, however, adopting the full-duplex relay into the CRNs can result in a series of problems. Due to the full-duplex operation at the CR, the primary receiver simultaneously receives interference from CS and CR. Therefore, in order to satisfy the interference constraint, the transmission powers at the CS and CR, must be lower than those in the half-duplex case.

Moreover, the signal receptions at the CR and NOMA-strong user are confronted with SI and co-channel interference, respectively. To deal with these challenges, we propose to employ attractive techniques, such as joint power allocation between the CS and CR and/or CR beamforming design (in case of MIMO-CR). Fig.~\ref{fig:3}(b) depicts the achievable rate region of the CRN in Fig.~\ref{fig:3}(a) for different levels of predetermined maximum tolerable interference level, $\Ith$, at the primary receiver, where CR is equipped with $\Ntx = 3$ transmit antennas and $\Nrx=2$ receive antennas. With the optimum scheme, receive and transmit beamformers at the CR and the transmit power levels at CS and CR are jointly optimized so that the $\CU$'s capacity is maximized, while the $\CUU$'s capacity is guaranteed to be above a certain value. With suboptimum design, the receive and transmit beamformers at the CR are designed using the MRC and MRT principles, respectively, and the transmit power levels at CS and CR are optimized. As can be observed, the achievable rate of the $\CU$ and the $\CUU$ increase significantly compared to the half-duplex CRN, particularly when the optimum scheme is used.

Full-duplex operation also offers the potential to achieve simultaneous sensing and transmission in cognitive radio NOMA networks (CR-NOMA). Specifically, a full-duplex cognitive transmitter (CS or CR) is able to dynamically sense the spectrum band and determine if the primary users are busy or idle, and at the same time decide to send data or keep silent. One possible approach to realize online spectrum sensing in CRNs is to use multiple antennas at the full-duplex cognitive transmitter, where some ($\Nsx$) antennas are allocated for sensing, some ($\Ntx$) antennas for transmitting data, and some ($\Nrx$) antennas for receiving data.
This approach, however, requires design of new signal processing techniques, resource allocation algorithms, and antenna selection rules.
To this end, new algorithms require to jointly take into account different quality-of-service levels for NOMA users.

\section{Resource Management in Full-duplex NOMA Networks}
Resource management is important for improving the system performance and at the same time to guarantee the quality-of-service (QoS) requirements. With full-duplex functions applied to the BS, resource management strategies for the downlink and uplink transmissions are closely related in a full-duplex system. This would generate higher computational complexity, compared with a half-duplex system where the user scheduling design for the downlink is typically independent of that for the uplink. The introduction of NOMA further aggravates the problem. Below we provide several insights into how the resource management problem in full-duplex NOMA networks can be addressed.

\begin{figure}[ht]
\centering
\hspace{-12em}
\includegraphics[width=130mm, height=100mm]{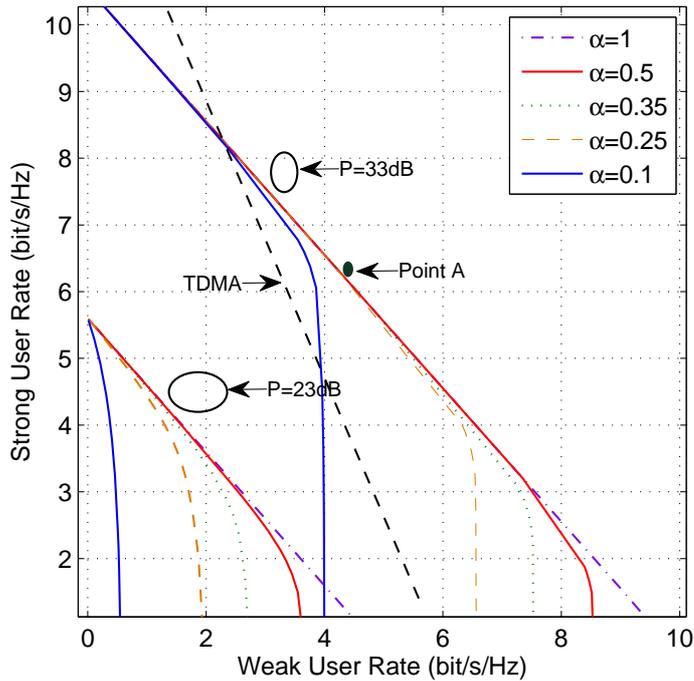}
\vspace{-0.5em}
\caption{Two-user rate region achieved by beamforming with superposition coding.}
\label{fig:Rate_Region}
\vspace{-1em}
\end{figure}

First, instead of merely considering time, frequency, and space as the basic elements for resource optimization, for NOMA transmission, it is important to further investigate whether it is beneficial for multiple users to share one `resource block'. Separating them at the receivers by implementing SIC, may lead to poorer throughput performance than orthogonal transmission techniques. The user scheduling policy must consider the uplink-downlink interference due to full-duplex operation so that the downlink user is not significantly interfered by the uplink signals and a good balance between downlink and uplink performance can be achieved. Fortunately, since the BSs usually transmit signals with stronger power than the uplink users, the uplink-downlink interference can be commonly treated as extra noise at the downlink users \cite{FD-NOMA} or in another word, it is unlikely to eliminate the uplink-downlink interference at the downlink receiver end through SIC.

To decouple these challenges brought by NOMA and full-duplex operation for resource management, let us first consider the impact of NOMA on downlink transmission. It is temporarily assumed that the user selections for the downlink and uplink are independent of each other so that the uplink-downlink interference is simply treated as extra noise at the receiver ends. A recent research \cite{Opt} studies the beamforming optimization algorithm for the downlink NOMA transmission, where the transmitter is equipped with multiple antennas and the user is equipped with single antenna. We use beamforming with superposition coding (SCBF) to denote such a beamforming process in NOMA networks. The algorithm we derived in \cite{Opt} is general and can be applied to communication scenarios such as cellular networks and satellite communications. The spatial correlations between the channels related to different users as well as the channel asymmetry, created for instance by the receiver heterogeneity, are the two major factors influencing the efficiency of SCBF. Specifically, in \cite{Opt} the two-user rate region is used to characterize the maximum throughput performance achieved by applying SCBF to the downlink transmission. With the spatial channel correlation fixed, Fig. \ref{fig:Rate_Region} shows how the two-user rate region is related to the transmission power (represented by different values of P) and to the degree of channel asymmetry (represented by different values of $\alpha$, which is the attenuation factor of the weaker channel with respect to the stronger one). The comparison between SCBF and time-division multiplexing (TDM) becomes straightforward, as the rate region achieved by TDM is approximately the line segment between the x- and y-axis intercepts of the SCBF rate region curve.

The extension of the result to cases with more than two downlink users can then be made by user grouping and by combining NOMA with orthogonal transmission methods, such as ZF and TDM, to minimize inter-group interference. Translating the uplink-downlink interference into the channel asymmetry, we can also conveniently use the above result to achieve a joint design for the downlink and uplink user scheduling. For example, when the operational point lies in Point A, as shown in Fig. \ref{fig:Rate_Region}, increasing the channel asymmetry, or equivalently introducing more uplink-downlink interference to the weaker user will not degrade the performance as long as the channel ratio $\alpha$ is above 0.25.

We note that the complexity of the link scheduling problem typically grows very rapidly with the number of users. Heuristic methods are often desired that achieve good compromise between the performance and computational complexity. Our analysis have so far assumed full spectrum sharing among all users. It is straightforward to generalize the result to the cases where flexible frequency allocation is enabled.
\section{Future Research Challenges }
In the following, we outline some interesting research challenges and future directions for full-duplex NOMA systems.

\textbf{Interference management for full-duplex NOMA systems:} Full-duplex operation creates a large number of simultaneous transmissions in multi-cell settings which can cause elevated interference levels in wireless systems. Unlike traditional full-duplex communications, interference will adversely effect a NOMA user's ability to successfully decode messages. Therefore, when full-duplex NOMA systems are deployed, interference must be carefully studied so that effective countermeasures can be designed. To this end, power control schemes, scheduling, and error control coding schemes could be investigated in detail. Moreover, employment of inter/intra-cell interference suppression techniques and approaches such as interference alignment with low complexity are promising as worthwhile directions to pursue.

\textbf{Low complexity multi-antenna schemes:} There has been a significant interest to develop low complex MIMO and massive MIMO systems. To this end, antenna selection is a popular technique. However, antenna selection used in traditional full-duplex systems can not be directly applied because the antenna selection in full-duplex NOMA systems must account both SI as well as user links into consideration to satisfy NOMA constraints. Hence design of antenna selection schemes for full-duplex NOMA systems that strike a good balance between performance and implementation complexity such as the feedback information of CSI is a promising research direction.

\textbf{User pairing in full-duplex settings:} The performance of the NOMA is very dependent on which users are selected to pair. Most existing user pairing methods in NOMA systems have been proposed to support two users. To exploit the benefit of the full-duplex operation in NOMA systems, however, general user pairing algorithms can be developed which are able to go beyond typically assumed dual-user pairing case. To this, the impact of SI and inter-user interference has to be carefully considered in new user-pairing designs.

\textbf{Massive machine-to-machine (M2M) communication access with energy harvesting:} NOMA is touted as a promising approach to implement massive access in emerging  M2M communication networks. However, devices in such networks suffer from energy constraint issues. To enable perpetual operation of M2M devices, energy harvesting techniques offer attractive solutions.  Adoption of such techniques in full-duplex NOMA systems require sophisticated power and time-split optimization solutions as compared to OMA transmissions. Further, in addition to transmit power, SI cancellation and SIC will increase the node level power consumption. Therefore, solutions based on intelligent power management algorithms, low-complexity optimization procedures and efficient hardware implementation are required to reap the benefits of energy harvesting full-duplex NOMA systems.

\section{Conclusion}
In this article, the concept of NOMA with full-duplex operation was discussed. We first reviewed the state-of-the-art and discussed uplink and downlink transmission, cooperative relay and cognitive radio with full-duplex NOMA operation. Further, the design of non-orthogonal beamforming with superposition coding was discussed as a natural way of extending NOMA to MIMO communications. Key research challenges and potential solutions that would hasten the practical deployment of full-duplex NOMA systems were also identified.
\bibliographystyle{IEEEtran}

\begin{thebibliography}{1}

\bibitem{Ashutosh:JSAC:2014}
A. Sabharwal, P. Schniter, D. Guo, D. W. Bliss, S. Rangarajan,
and R. Wichman, ``In-band full-duplex wireless: Challenges and
opportunities," \emph{IEEE J. Sel. Areas Commun.,} vol. 32, pp. 1637-1652, Sep. 2014.

\bibitem{Riihonen:JSP:2011}
T. Riihonen, S. Werner, and R. Wichman, ``Mitigation of loopback self-interference in full-duplex MIMO relays,'' \emph{ IEEE Trans. Signal Process.}, vol. 59, pp. 5983-5993, Dec. 2011.

\bibitem{Sabharwal:TWC2012}
M. Duarte, C. Dick, and A. Sabharwal, ``Experiment-driven characterization
of full-duplex wireless systems," \emph{IEEE Trans. Wireless
Commun.,} vol. 11, pp. 4296-4307, Dec. 2012.

\bibitem{Saito:VTC2013}
Y. Saito, Y. Kishiyama, A. Benjebbour, T. Nakamura, A. Li, and
K. Higuchi, ``Non-orthogonal multiple access (NOMA) for cellular
future radio access," in \emph{Proc. IEEE 77th Veh. Technol. Conf. (VTC'13),}
Dresden, Germany, June 2013, pp. 1-5.

\bibitem{Ding:Survay}
Z. Ding, X. Lei, G. K. Karagiannidis, R. Schober, J. Yuan, and
V. K. Bhargava, ``A survey on non-orthogonal multiple access for
5G networks: Research challenges and future trends," \emph{IEEE J. Sel.
Areas Commun.}, vol. 35, pp. 2181-2195, Oct. 2017.

\bibitem{Ding:TVT:2016}
Z. Ding, P. Fan, and H. V. Poor, ``Impact of user pairing on 5G nonorthogonal multiple-access downlink transmissions," \emph{
IEEE Trans. Veh. Technol.}, vol. 65, pp. 6010-6023, Aug. 2016.

\bibitem{Zhiguo-MIMO}
Z. Ding, R. Schober, and H. V. Poor, ``A general MIMO framework for NOMA downlink and uplink transmission based on
signal alignment," \emph{IEEE Trans. Wireless Commun.}, vol. 15, pp. 4438–4454, June 2016.

\bibitem{FD-NOMA}
Y. Sun, D. W. K. Ng, Z. Ding, and R. Schober, ``Optimal joint power
and subcarrier allocation for full-duplex multicarrier non-orthogonal
multiple access systems," \emph{IEEE Trans. Commun.,} vol. 65, pp. 1077-1091, Mar. 2017.


\bibitem{Z.Zhang}
Z. Zhang, Z. Ma, M. Xiao, Z. Ding, and P. Fan, ``Full-duplex device-to-device aided cooperative non-orthogonal multiple
access," \emph{IEEE Trans. Veh. Technol.}, vol. 66, pp. 4467-4471, May 2017.


\bibitem{Nallanathan:FDNOMA}
X. Yue, Y. Liu, S. Kang, A. Nallanathan, and Z. Ding, ``Outage
performance of full/half-duplex user relaying in NOMA systems," in
\emph{Proc. IEEE Intl. Conf. Commun. (ICC'17),} Paris, France, May 2017,
PP. 1-6.

\bibitem{Caijun:CLET:2016}
C. Zhong and Z. Zhang, ``Non-orthogonal multiple access with
cooperative full-duplex relaying," \emph{IEEE Commun. Lett.,} vol. 20, pp.
2478-2481, Dec. 2016.

\bibitem{Jungho:CLET:2016}
J. So and Y. Sung, ``Improving non-orthogonal multiple access by forming relaying broadcast channels," \emph{ IEEE Commun.
Lett.}, vol. 20, pp. 1816-1819, Sept. 2016.



\bibitem{Z.Ding}
Z. Ding, M. Peng, and H. V. Poor, ``Cooperative non-orthogonal multiple access in 5G systems," \emph{IEEE Commun. Lett.},
vol. 19, pp. 1462-1465, Aug. 2015.


\bibitem{J.Kim}
J. Kim and I. Lee, ``Non-orthogonal multiple access in coordinated direct and relay transmission," \emph{ IEEE Commun. Lett.},
vol. 19, pp. 2037-2040, Nov. 2015.



\bibitem{Opt}
X. Shi, J. S. Thompson, M. Safari, and R. Liu, ``Beamforming with superposition coding in multiple antenna satellite
communications," in \emph{ Proc. IEEE Intl. Conf. Commun. (ICC'17), (Workshop on Satellite Communications)}, Paris, France,
May 2017, pp. 705-710.

\end{thebibliography}

\end{document}